# Diverse Perspectives Can Mitigate Political Bias in Crowdsourced Content Moderation

JACOB THEBAULT-SPIEKER, Information School, University of Wisconsin – Madison*, USA

SUKRIT VENKATAGIRI, Center for an Informed Public, University of Washington*, USA

NAOMI MINE, Information School, University of Wisconsin – Madison, USA

KURT LUTHER, Department of Computer Science, Virginia Tech, USA

In recent years, social media companies have grappled with defining and enforcing content moderation policies surrounding political content on their platforms, due in part to concerns about political bias, disinformation, and polarization. These policies have taken many forms, including disallowing political advertising, limiting the reach of political topics, fact-checking political claims, and enabling users to hide political content altogether. However, implementing these policies requires human judgement to label political content, and it is unclear how well human labelers perform at this task, or whether biases affect this process. Therefore, in this study we experimentally evaluate the feasibility and practicality of using crowd workers to identify political content, and we uncover biases that make it difficult to identify this content. Our results problematize crowds composed of seemingly interchangeable workers, and provide preliminary evidence that aggregating judgements from heterogeneous workers may help mitigate political biases. In light of these findings, we identify strategies to achieving fairer labeling outcomes, while also better supporting crowd workers at this task and potentially mitigating biases.

# 1 INTRODUCTION

In recent years, social media companies have begun to focus on the risks associated with political messaging on their platforms, including issues of fairness — or bias — and disinformation. The 2016 US presidential election, and an associated disinformation campaign run by Russia's Internet Research Agency (RU-IRA), was a catalyzing event that led social media companies to grapple with the political content on their platforms, study the role it plays in shaping public opinion, and begin thinking about imposing limitations [63]. For instance, in 2019, Twitter's CEO, Jack Dorsey, announced that Twitter would ban all political advertising [62] and Spotify followed suit in 2020 [43]. In 2020, Facebook enacted a policy in which accounts posting political ads needed to go through an authorization process and label all political ads [37], and Nextdoor implemented policies disallowing conversations about national political campaigns in the US [50]. Google and YouTube instituted much coarser targeting controls for political advertising [23] in 2022, and LinkedIn deployed a feature allowing users to hide political content in their feeds [5].

---

* Work was partially completed while this author was affiliated with Virginia Tech.

Beyond issues of disinformation, social media companies face growing concerns about perceived political biases on their platforms. One prominent example of these concerns was when Facebook decided to automatically curate, and then subsequently shutter, their Trending News feature due to critiques of political bias [18]. More recently, many states in the US have begun to explore legal avenues attempting to guarantee "fair" content moderation practices [10]. For instance, in Texas, a law and subsequent court cases seek to ensure that a user of a social media system cannot be "censored because of their viewpoint", predicated on the perception that some users are being censored because of their viewpoint. Similarly, Florida recently passed a law making it illegal to ban politicians' accounts [46].

Enforcement of political content policies, and concerns about (un)fair treatment of some political groups, creates an environment in which social media companies need to decide which content is political, and whether or not it should be allowed on their platform — in other words, to perform political content labeling and moderation. To achieve this efficiently and at scale, platforms often rely on algorithmic content labeling and moderation techniques. However, automated approaches have been shown to be insufficient when the content is highly subjective and contextual (e.g., hate speech or disinformation) [18]. Thus, human content moderators — who have been shown to evaluate subjective, contextualized information more effectively — are often hired to augment or help train algorithmic systems [24].

However, incorporating human labor into these systems brings its own risks: inefficacy and systematic biases in judgement. Even for experienced human moderators, identifying and labelling subjective content remains challenging [6], and, particularly in a politically polarized social space, human content moderators may make unfair or biased decisions. Of particular concern to crowd labeling and evaluation is the risk of biases stemming from the composition of crowds [14], i.e., who comprises the crowd. In organizational psychology, prior work suggests that heterogeneous teams can be more effective at achieving their stated goals because a more diverse set of perspectives are included, but in other settings can actually hinder a teams' effectiveness [30]. According to Duan et al. [20], it is unclear how the heterogeneity of crowd makeup might play out in political content labeling and moderation, as both approaches to aggregating crowd responses [36] and crowd composition [58] are avenues of possible bias.

Given the public and legal interest in issues of social media and politics, and the potential for unfair content labeling and moderation, crowds' effectiveness and fairly labeling political content becomes a high-stakes focal area for research. In our work here, we directly focus on this question: *how effective are human crowds at identifying and labeling political content on social media, without producing unfair outcomes?*

In this paper, we experimentally evaluate this question by measuring crowd workers' success against a ground-truth set of political social media posts, in a US context, using two different task interfaces: (1) a naive baseline and (2) an industry policy around political content moderation. Our findings make four primary contributions:

(1) Our results problematize treating crowd workers as individually interchangeable because workers' political perspectives can harm crowds' effectiveness at identifying political content. Techniques that aggregate across workers' decisions, but account for a diversity of perspectives, perform best in our study.
(2) Further, we find that in some cases, crowds can create systematically unfair outcomes in how well some kinds political content gets labeled. However, our results suggest that aggregation techniques which increase the diversity of perspectives may help ameliorate these biases and their associated risks.
(3) We also show that it may be important for platforms to be opinionated in how policies are specified through interface design, though this finding may vary with crowd workers' political orientations.
(4) Finally, we develop implications for social media users, researchers, and industry practitioners.

## 2 RELATED WORK

In recent years, political content online and in social media has been the subject of intense public discussion, as well as an active area of research. Our work here builds on and contributes to three bodies of research in this space: (1) political content on social media, (2) unfairness in content moderation, and (3) social biases in other distributed work settings.

### 2.1 Recognizing Political Content Online

Political content, and political but untrue or misleading topics in the news, have been the focus of recent research that focuses in online information ecosystems. For instance, Pennycook and Rand [51] showed that higher credibility of news sources leads people to trust those sources more; people across the political spectrum found mainstream news sources more trustworthy than partisan outlets. More recently, Pennycook and Rand [52] showed that people's ability to discern partisan "fake news" from real news is not partisan, but instead relates to analytical thinking about the news headlines.

Beyond evaluating news topics themselves, political content on social media has also become an important focal area, particularly given state actors' weaponization of politics on social media in recent years. For example, Starbird et al. [63], through a series of case studies, showed how targeted disinformation campaigns are a collaborative and participatory phenomenon that are built on top of and intertwined with social computing systems. Others [18, 19, 49, 70] have explored various kinds of disinformation behaviors on social media. Examples of this work include: identifying hoaxes on Wikipedia [39], exploring approaches to identifying organized groups of "bots" perpetrating disinformation on social media [1], and characterizing the disinformation techniques used by governmental actors [42, 45]. More recently, Atreja et al. [7] explored what social media users want platforms to do in response to this dis- and misinformation.

Overall, it is unclear how political orientation of participants should be leveraged for the purposes of design, because results are mixed in in many of these recent studies [7, 51, 52]. Pennycook and Rand [51] found that liberals with higher cognitive reflection were more effective at discerning credible news sources. Pennycook and Rand [52] extend this finding and show that people are able to more effectively discern real news from fake news when the headlines align with their own political orientations. However, Atreja et al. [7] found that participants preferred platforms take more action against potentially misleading content from the opposite political perspective. Our research here adds further evidence to this topic, showing how participants' political views impact performance in a content moderation setting, and potential bias mitigation techniques platforms may adopt.

### 2.2 Content Moderation Biases in Social Media

User-generated content moderation has been a focus of social computing research for over 25 years [12], and many researchers have studied the general effects that content moderation has on social media communities [16, 41, 47]. There are a variety of reasons organizations moderate content, including setting norms [26, 41], mitigating legal risks [10], and protecting users from harmful content [15]. However, the scale and breadth of social media has created contexts that push the limits of how effective automated content moderation approaches can be. This has led large social media companies to re-incorporate humans into their previously fully-automated content moderation processes [29]. However, human content moderation is difficult for topics that are subjective or complex [6], and techniques that seek to take advantage of the "wisdom of the crowd" are only moderately effective [11, 53, 66, 69]. The difficulty of content moderation in contextual settings can also lead to disproportionate amounts of content being removed for some groups [25].

Biases in social media has become an important topic in the public sphere. For instance, one poll in 2018 suggested that the majority of the US thinks that social media companies exhibit biases against politically conservative content [59]. This controversy has led to large companies changing [21] and eventually removing [37] major software features. Other accusations of bias in political content moderation include the removal of a Moroccan secularist group from Facebook [70], and allegations that Facebook applies hate speech rules differently between Palestinian and Israeli content [49], among others [18]. Ma and Kou [44] explored how social media users perceive issues of bias.

More recently, and most closely tied to our research here, researchers have turned their focus to issues of political unfairness in content distribution and moderation. Chowdhury and Belli [17] found that Twitter's algorithm (at the time) was disproportionately amplifying conservative political content. Sang et al. [55] systematically found different levels of negative and toxic content depending on the political valence of the topic, echoing prior work [35, 60]. Hu et al. [32], focusing on possible solutions, found that when structuring moderation processes as a jury, with an included deliberation period, decisions remain consistent across trials. Researchers in the FAccT community recognize the risks of these kinds of human judgement and labelling biases, and have begun exploring ways to address the potential for discriminatory outcomes that comes from data labeling [9].

### 2.3 Social Biases in Distributed Work

Researchers in the field of Computer-Supported Collaborative Work (CSCW) have also studied social biases in distributed work settings other than content moderation. Kuo et al. [40] explored the feasibility of crowds in identifying news sources' political valence, using similar techniques to our work here and prior work by Thebault-Spieker et al. [65]. Most closely related to our work here is the study by Hube et al. [33] evaluating the potential for biases in natural language subjective labelling tasks performed on Amazon Mechanical Turk. They found that participants with stronger opinions were more likely to exhibit biases in these tasks. Hube et al. also explored three bias mitigation strategies: a Bayesian truth serum approach wherein participants specify what they think others will say, a reminder about the potential for bias, and personalized nudges. Our work differs in two key ways: (1) we focus on adversarial political content generated as a part of a disinformation campaign, and (2) we focus explicitly on the feasibility of the content moderation task. More recently, Duan et al. [20] explored the impact that political diversity has on biases in crowdwork settings, and found that exposure to diverse perspectives may have the potential to help mitigate some of these biases.

### 2.4 Our Work Here

Taken together, these bodies of work all point to the importance of understanding the interaction between political content, biases in social media systems and in political content moderation, and the potential for crowd workers to be able to identify and label political content. However, none of these studies focuses on this intersection itself. Our work here directly addresses this gap in the literature by experimentally investigating the feasibility of crowds labeling political content, particularly focusing on the potential for unfair outcomes for some kinds of political content.

## 3 METHODS

Using crowds to label subjective content like politics can be difficult, and risks creating biases that advantage one perspective over another [58]. Therefore, we experimentally study how effective crowd workers are at labeling political content. Specifically, our research questions are:

- **RQ1a:** How well can humans recognize and label political content on social media?

- **RQ1b:** Are there systemic variations in crowd workers' ability to effectively label political social media content?
- **RQ2a:** Do crowdsourcing aggregation techniques improve the feasibility of content moderation to identify political content on social media?
- **RQ2b:** How does heterogeneity of team perspectives influence the effectiveness of crowd aggregation techniques?

### 3.1 Stimulus Set

To address these questions, we required a dataset of social media posts that fulfilled two criteria. First, the dataset must be representative of real-world political content on social media. Second, it must have robust ground-truth labels describing its political content. We found no pre-existing datasets that satisfied both criteria, so we decided to generate one.

After an extensive search, we started with the RU-IRA Twitter dataset built by Linvill and Warren [42], because Linvill and Warren had already produced some relevant labels for the content as part of their own analysis, and because the context around this data was very clearly political: Twitter generated a list of 2,848 accounts that they had identified as part of the Russian IRA disinformation campaign in 2016, and shared this list with the US House Intelligence Committee [57]. Linvill and Warren used a subset of this data (posts sent between June 18, 2015 and December 31, 2017) to develop an understanding of the behavior of these accounts, and a taxonomy for labelling the posts by the type of account that sent the post.

To build our stimulus set, we started with Linvill and Warren's initial dataset consisting of nearly 3 million posts, and we selected all posts made by three types of accounts defined by Linvill and Warren: *LeftTroll*, *RightTroll*, and *NewsFeed*. This produced a set of posts made by political accounts (LeftTroll and RightTroll) and non-political accounts (NewsFeed), based on definitions provided in the original publication [42]. We then randomly sampled 1500 posts from each category, for a total of 4,500 tweets. Each post was labelled based on the kind of account it came from (i.e., RightTroll LeftTroll, or NewsFeed), providing some initial information about the political-valence of the account sending the post. While the labels Linvill and Warren used made sense for their analysis, our purpose here is different, so we refer to RightTroll, LeftTroll, and NewsFeed accounts as "conservative-targeting", "liberal-targeting", and "news-sharing", respectively.

*3.1.1 Developing ground truth labels.* Linvill and Warren note that their categorizations describe overall account behavior; however, subsequent work by Linvill and Warren [42] showed not all posts created by these accounts were necessarily political in nature, due to the audience-building efforts made by these accounts prior to posting political content. Therefore, we also needed to develop a set of ground truth labels about whether the posts were political or not. To do so, we employed a definition of "political" content from political communication theory [48].

Two expert members of our team who had research experience on adversarial crowds and industry experience with content moderation on political content coded these the posts in our dataset and developed a set of ground truth labels. They first coded 10% of our dataset and reached approximately 76% agreement. They met to resolve disagreements, and then coded an additional 10% of our dataset. This second round of coding reached approximately 91% agreement, and Cohen's Kappa of 0.81, indicating nearly perfect agreement. We then split the remaining 80% of the dataset in half, and each coder then independently coded 40% of the dataset. Overall, our resulting ground-truth labels from this coding process found that political and non-political posts are relatively evenly distributed across our whole dataset — our coders labeled 38% of the posts in our data as political. The 38% of political posts in our overall dataset are composed of approximately 4% of posts from liberal-targeting accounts, approximately 17% of posts from conservative-targeting accounts, and approximately 16% of posts from news-sharing accounts.

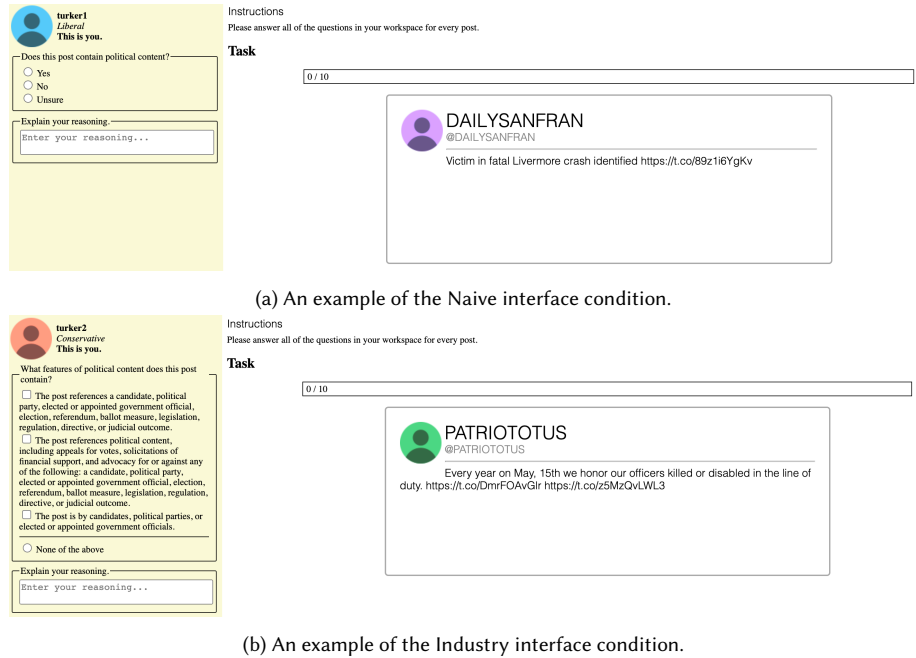

(a) An example of the Naive interface condition.

(b) An example of the Industry interface condition.

Fig. 1. Screenshots of both interface conditions. Participants only saw one of these interfaces across all 10 posts.

### 3.2 Experimental Setup

Fundamentally, our work focuses on the feasibility of using human content moderators to identify political posts on social media. After all, human decisions influence multiple aspects of the content moderation process, from building initial training data for machine learning models to human-in-the-loop hybrid approaches [29]. However, prior work has shown that task (i.e., user interface) design can substantially impact people's task performance [3, 22]. Furthermore, as large social media platforms began to institute policies banning political content, critics raised a fundamental question: what is considered "political" content? Successfully enforcing these policies requires a specific definition or rubric.

Therefore, we developed two alternative task designs, each of which operationalizes a different definition of "political." In our Naive condition (Figure 1a), participants answer the question, "Is this political?" This condition provides a Naive baseline comparison against our second condition, and also captures what kinds of content people perceive as political without other definitions being provided. In the Industry condition (Figure 1b), participants apply Twitter's official policy [13] to the posts they are evaluating. This condition provides a metric of how effective real-world guidelines from a major social media platform are.

Participants were shown a total of 10 posts: four unique, randomly selected, left- and right-valenced posts (two each from the liberal-targeting and conservative-targeting accounts), four unique, randomly selected, neutrally-valenced posts (news-sharing), and a duplicate of one political and one non-political post each. To address RQ2 and enable post-hoc analysis of crowd aggregation techniques, we ensured that each set of 10 posts was seen by multiple participants. Each post was assigned an 'aggregation team', composed of three random Amazon Mechanical Turk workers who accepted our task. As workers started the task, they were assigned to a consistent interface condition, but the composition of the team was treated as a random variable (i.e., was not controlled).

In this between-subjects design, we also worked to address two additional methodological challenges. We needed to ensure that (1) responses would reflect real content moderation behavior, and (2) participants would be most likely to demonstrate biases that otherwise might not have been revealed due to subject expectancy bias [54]. Therefore, we employed an IRB-approved deception study in which we told participants they were moderating content for a real website that was banning all political content. This deception was designed to facilitate realistic content moderation behavior, including naturally-arising biases.

We implemented the crowdsourced moderation system as a Python/Flask-based web application. The left side of the interface displayed (from top to bottom) the worker's information (Mechanical Turk ID, political orientation, and color-coded silhouette avatar), the moderation interface (with checkboxes and/or radio buttons for the Naive or Industry condition), and a textbox where the worker was asked to optionally "explain your reasoning." The right side of the interface displayed the task instructions, a progress bar showing the number of completed tasks (out of 10), and the current tweet requiring a moderation decision. The system displayed the username of the tweet poster and the tweet content (including any links) on a simulated nondescript social media page.

### 3.3 Recruitment and Procedure

We recruited all participants from Mechanical Turk. We required that participants be residents of the United States to ensure that they had relevant political context for the posts, but imposed no other qualifications. Each participant was paid at least $1.21 for 15 minutes of work, or $7.25/hr. We asked participants to self-report their own political orientation at the beginning of the task in two ways: by political orientation and by party affiliation. With regard to political orientation, we asked participants to enter their political orientation, and they were given three options: liberal, conservative, or other (with an open text field). With regard to party affiliation, we asked participants which of the following best described them, and they were given three options: Democrat, Republican, or Other (with an open text field). In the results below, we identify these self-reported answers as 'Democrat', 'Republican', or 'Other' respectively. Participants then completed the content moderation task in one of the two conditions above.

After completing the moderation tasks, we presented participants with debrief information about our deception study, and gave them the option to consent at that time, consistent with IRB recommendations. Participants were paid regardless of their choice to consent, and those who did not consent were excluded from data analysis. Beyond participants who did not consent, we also excluded participants who were inconsistent in how they answered the duplicated posts, or if they selected ['conservative' and 'Democrat'] or ['liberal' and 'Republican'] in the initial self-report questions. While the latter are legitimate responses, they represent too small of a minority group in our dataset to enable robust analysis, particularly with regard to aggregation.

We hired a total of 627 Mechanical Turk workers, 190 of whom did not consent to participate in our deception study, and were therefore excluded. Another 43 participants were inconsistent in their answers on the duplicate posts, and were excluded as well. Finally, another 113 were inconsistent in their political orientation and affiliation. We ended up with 468 participants who consented to participate and were consistent in how they answered the duplicated questions. These 468 participants ultimately labelled 1,149 posts that were randomly sampled from our original 4,500-post stimulus set.

*3.3.1 Metrics.* Because our study focuses on the ability for moderators to accurately assess and label political content, we evaluate our results using two metrics: precision and recall. These metrics reflect common usage in academic research and industry practice for measuring content moderation performance [61]. *Precision* describes the percent of posts labelled as political that were correct (high precision indicates relatively few false positives), and *recall* describes the percent of actually political posts that were correctly labelled (high recall indicates relatively few false negatives). As a baseline, if one were to randomly assign whether a post was political or not based on a coin flip, across 10 posts we would expect a precision of 0.6, and a recall of 0.5. In some cases, we also present results in terms of accuracy, for better interpretability.

## 3.4 Methodological Limitations

Our methodological decisions impose some limits to the generalizability of our work that we view as opportunities for future work. First, our study is focused on the United States. We required our moderators be from the US and we used posts that were relevant to US politics, which may limit the applicability of our findings in other cultural or geographic settings. Second, our work here focused on a data set of political content created as a part of an active disinformation campaign that potentially reached millions of people [31]. It is likely this content looks similar to more general political content posted by "real" users on social media platforms; after all, the intent of disinformation is to mislead and deceive. However, it is not clear to what extent our results generalize to this broader context. Relatedly, we defined "political", for the sake of our study, based on an expert definition from the field of Political Communication. This operational definition may exclude some political content, putting some constraints on the generalizability of our definition.

Finally, our study recruited crowd workers to complete only 10 moderation tasks each. We did not place qualifications on the amount of moderation experience workers had, or collect self-reported data on their moderation experience; thus, the moderation experience of the average worker in our study could vary. While our study design replicates a common real-world scenario — in which companies hire on-demand workers for small batches of content moderation microtasks with little or knowledge of the workers' skills or experiences [24] — it is possible that moderation experience may affect worker performance and either close or widen some of the group differences described above.

## 4 RESULTS

|  | **Non-Political** | **Political** |
| --- | --- | --- |
| *news-sharing* | @onlinememphis: 'Tremendous' traffic impact expected after bridge collapse https://t.co/hMPr8OHSzb https://t.co/PbgIwUVIUN | @dailysanfran: Ryan tells GOP there's agreement on tax and spending bill #politics |
| *liberal-targeting* | @policestateme: https://t.co/0RxxAhvXeD | @blacknewsoutlet: Chicago police tries to round up #Chicago protesters. #BlackLivesMatter #BlackTwitter #LaquanMcDonald https://t.co/kY0XyvA3Ng |
| *conservative-targeting* | @cameericlaar: Kathy Griffin Praises Kaepernick for His 'Activism' https://t.co/9Uhqds0aBd | @debesstrs: RT kinni00: HANNITY Responds to McMaster Giving Clearance to Susan Rice: ""I'd Like an Explanation from White House… https://t.co/1Z3DnFlJRS |

Table 1. Representative example posts from our data set, each post is shown in the format @username: tweet content

Before describing our results for each research question, we first present a table of example posts in our data set (Table 1). We show one example of political and non-political posts across each type of content (liberal-targeting, conservative-targeting, news-sharing), for a total of six posts. We ensured these posts were representative by selecting posts that had the median number of people label it correctly, for each group.

### 4.1 Individual Content Moderation Results

*4.1.1 RQ1a: How well can humans recognize and label political content?* For our first research question, we evaluate how individual participants performed in each condition. The mean precision and recall were 0.77 (std. dev=0.28) and 0.72 (std. dev=0.31), compared to random baselines of 0.6 and 0.5, respectively.

| Precision | Recall |          | Precision |      |      | Recall |      |      |
|-----------|--------|----------|-----------|------|------|--------|------|------|
|           |        |          | Dem.      | Oth. | Rep. | Dem.   | Oth. | Rep. |
| 0.73      | 0.69   | *Naive*  | 0.72      | 0.63 | 0.81 | 0.69   | 0.76 | 0.68 |
| 0.79      | 0.67   | *Industry* | 0.78    | 0.81 | 0.79 | 0.68   | 0.69 | 0.63 |

Table 2. *Left*: By experimental condition, *Right*: further by participant political orientation.

However, focusing only on the mean masks key differences between interface conditions. Breaking the results down by interface (Table 2), we see that the our Industry policy condition achieves a slightly higher precision of 0.79, while the Naive condition reaches 0.73. Both precision and recall outperform a random baseline (0.6 and 0.5 respectively).

In a follow-up analysis, we considered whether people with specific political viewpoints perform better in some interface conditions than others. Results in Table 2 are mixed: self-identified Republicans perform better than other groups in the Naive condition, though the Industry condition helps decrease that gap. We do not see evidence that the Industry condition raises the precision for Republican participants, who show precision of 0.79 for Industry and 0.81 for Naive conditions, but rather that the Industry condition enables other groups (Democrat and Other participants) to reach precision levels of 0.78 and 0.81, comparable to Republicans.

|          | liberal-targeting | news-sharing | conservative-targeting |
|----------|-------------------|--------------|------------------------|
| *Naive*    | 69%               | 78%          | 83%                    |
| *Industry* | 73%               | 76%          | 84%                    |

Table 3. Accuracy rates, broken down by the type of content and the experimental condition in which it was shown

*4.1.2 RQ1b: Are there systemic variations in crowd workers' ability to label political content?* Our review of prior work suggests that it is possible that crowds judge some kinds of content systematically differently (e.g. [14, 52]), giving rise to biases in the content moderation process that risk unfair content moderation outcomes. Indeed, this may be an underlying cause of the effectiveness issues we identify in the previous section.

Turning to how these labeling dynamics affect fairness across types of political content, our results in Table 3 show that liberal-targeting content is substantially less accurately labelled than conservative-targeting content, regardless of the interface condition. Where liberal-targeting label accuracy ranges from 69% to 73% across conditions, conservative-targeting label accuracy is much more consistent across conditions, between 83% and 84%. The difference between accuracy on conservative-targeting content and other types of content is as high as 14% in some cases.

These results also provide further evidence that the Industry condition may not always support more effective labeling for political content versus the Naive condition. The Industry condition achieves 73% and 84% accuracy for liberal-targeting and conservative-targeting content, respectively. By contrast, for news-sharing content, the Naive condition is most accurate with 78% accuracy, whereas the Industry condition is slightly smaller with 76% accuracy.

*4.1.3 Statistically validating these trends.* Accuracy, precision, and recall do not provide statistical confidence in these patterns, so to further explore how statistically confident we should be in these trends, we constructed a mixed effects logistic regression. Our dependent variable was whether a post was correctly labelled as "political" or not. This model specification also allows us to control for the random effect of each participant having seen multiple posts, while also modeling the impact of our variables (interface condition and post type) on the likelihood of being correct. To reflect how we break down our results above, we also included interaction variables between condition and post type.

Because this is a logistic regression and our independent variables are categorical, we needed to decide which categories would serve as the reference group, i.e., the intercept coefficient. Our results above suggest that the Industry condition differs meaningfully from the Naive condition for some groups of participants. Thus, we selected the Naive condition as the reference category for our interface condition variable. For our content type variable, we use news-sharing content as the reference category, allowing us to compare news-sharing content to liberal-targeting and conservative-targeting content. We also include an interaction term between our interface condition and our content type condition, based on our results in Table 3. Finally, for our political affiliation variable, we use participants with political affiliation of Other as our reference category, allowing us to compare the effects of being a Republican or Democrat against that baseline. To summarize, we use Naive-News-Sharing-Other as our reference category, so we can evaluate the differences between the Naive-News-Sharing-Other condition and the other permutations of our interface condition, political affiliation, and content type variables. All coefficients and p-values describe what is predicted to happen when the interface varies from the Naive condition, the content varies from news-sharing, and a participant's political affiliation varies from Other, or permutations thereof.

Table 4 both confirms and adds nuance to the trends described above. We see statistically significant differences between the reference news-sharing content and conservative-targeting and liberal-targeting content types. We also see a suggestive positive effect for the intersection between the Industry interface and liberal-targeting type content, providing additional statistical evidence for the trends we saw in Table 3. We also find that individual political affiliation does not have a statistically significant effect on individual likelihood of making a correct decision.

Table 4

|  | correct_decision |
|---|---|
| Constant | **1.467**** (0.162) |
| interface[*industry*] | −0.120 (0.160) |
| content[*liberal-targeting*] | **−0.555**** (0.145) |
| content[*conservative-targeting*] | **0.333*** (0.167) |
| party[*Republican*] | −0.151 (0.225) |
| party[*Democrat*] | −0.193 (0.203) |
| interface[*industry*] x content[*liberal-targeting*] | *0.394 x* (0.209) |
| interface[*industry*] x content[*conservative-targeting*] | 0.250 (0.235) |
| *Note:* | . p<0.1; *p<0.05; **p<0.01 |

### 4.2 Crowd Aggregation Results

*4.2.1 RQ2a: Do crowdsourcing aggregation techniques improve the feasibility of content moderation to identify political content?* While our individual-level results indicate moderate success at labeling political content, we also see evidence of potential for unfair outcomes between liberal-targeting vs. conservative-targeting content. These variations in effectiveness may create unfair outcomes for some political groups, which would provide evidence for common concerns about political bias on social media. In crowdsourcing, a common method to increase label quality is aggregating decisions from multiple workers. The most common aggregation model is *majority vote* [e.g. 68], in which a decision is only accepted if the majority of crowd workers in the group agree. However, some prior work [36, 38, 67] has begun to show that crowdsourced data may be able to overcome accuracy concerns by elevating minority perspectives. An alternative aggregation model to majority vote that accounts for minority perspectives is *one-yes* aggregation [38], in which a label is accepted when at least one of the crowd workers in the group applies it. Given our results above, we now turn our focus to evaluating these aggregation techniques, to improve the quality of political content labeling.

|  | *One-Yes* | | *Majority-Vote* | |
| --- | --- | --- | --- | --- |
|  | Precision | Recall | Precision | Recall |
| *Naive* | 0.88 | 0.41 | 0.75 | 0.41 |
| *Industry* | 0.91 | 0.35 | 0.73 | 0.36 |

Table 5. Precision and recall rates for both one-yes and majority-vote crowd aggregation models, by the experimental condition.

*4.2.2 Aggregation models: majority-vote vs. one-yes.* Table 5 shows two main trends in the results. First, the majority-vote aggregation approach is not meaningfully more effective in terms of precision than individual evaluators. Majority-vote aggregation shows precision of 0.75 in the Naive condition and 0.73 in the industry condition, versus 0.73 and 0.79 precision rates for individual evaluators. Further, majority-vote recall (0.41 in the Naive condition and 0.36 in the Industry condition) is much worse in comparison to both individual evaluators types (0.69 in the Naive condition and 0.67 in the Industry condition). By contrast, the one-yes aggregation model increases precision rates over individual workers by as much as 0.15 in some conditions, though recall declines to performing worse than a random baseline.

The second main trend also seems to reduce the difference between the Naive interface condition and the Industry interface condition we saw in the individual results. In the Naive condition, one-yes precision reaches 0.88, and the Industry condition has a one-yes precision score of 0.91.

Because the majority vote aggregation approach does not outperform individual workers in our data, all of our subsequent analyses focus on results using the one-yes aggregation model. Further, because one-yes aggregation can be effective so long as a single crowd worker uses a label, we also include aggregation teams with only two participants due to variations in participant hiring and inclusion criteria.

*4.2.3 RQ2b: How does heterogeneity of team perspectives influence the effectiveness of crowd aggregation techniques?* Because one-yes aggregation consists of multiple participants evaluating the same post, it does not make sense to break down precision and recall by individual political orientation. Instead, in Table 6, we show precision and recall rates for each interface condition, broken out by the composition of the aggregation team. Each category of team is labelled by whether the team includes Democrats (*Dem*), Republicans (*Rep*), or Others (*Oth*). For instance, a row labelled *Rep* represents aggregation teams composed entirely of Republicans, whereas a row labelled *Dem/Rep/Oth* represents aggregation teams composed of Democrats, Republicans, and Others.

Examining Table 6 in more detail, we again see that the Industry interface condition helps increase precision for teams composed of only Democrats, whereas the Naive interface condition shows higher precision for teams composed of Republicans. Beyond exclusively Republican teams however, in general we also see that the Industry interface condition enables increases in precision for some team compositions, and performs comparably for many others.

|  |  | Precision | Recall |
|---|---|---|---|
| **Naive** | *Dem.* | 0.85 | 0.41 |
|  | *Dem./Oth.* | 0.80 | 0.36 |
|  | *Oth.* | N/A | N/A |
|  | *Rep.* | 1.00 | 0.50 |
|  | *Rep./Dem.* | 0.96 | 0.40 |
|  | *Rep./Dem./Oth.* | 0.87 | 0.44 |
|  | *Rep./Oth.* | 1.00 | 0.53 |
| **Industry** | *Dem.* | 0.89 | 0.39 |
|  | *Dem./Oth.* | 1.00 | 0.30 |
|  | *Oth.* | 0.71 | 0.21 |
|  | *Rep.* | 0.75 | 0.30 |
|  | *Rep./Dem.* | 0.95 | 0.36 |
|  | *Rep./Dem./Oth.* | 0.83 | 0.29 |
|  | *Rep./Oth.* | 1.00 | 0.46 |

Table 6. One-yes aggregation accuracy rates, by team composition and and experimental condition.

However, the team composition analysis reveals another important trend. Teams that include heterogeneous composition (e.g., both Democrats and Republicans; or Democrats, Republicans, and Other participants) show meaningfully better precision than most homogeneously composed teams (e.g., exclusive Democrats). While teams composed of exclusively Republicans achieve perfect precision in the Naive interface condition, heterogeneously composed teams perform comparably well in both the Naive condition. In the Industry condition, both homogeneous Democrat (precision 0.89) and homogeneous Republican (precision 0.75) teams do not perform as well as heterogeneous Democrat-Republican compositions (precision 0.95) or teams that include Other members as well (precision ranging from 0.83–1.0).

Notably, across the board, recall (measuring the extent to which content should have been labeled as political, but was not) is fairly low for all team compositions, reflecting a general trend of aggregation techniques making recall worse in our data. In other words, our results suggest that aggregation can help decrease the rate of incorrectly identifying content as political, but is less successful at identifying the full set of the political content.

*4.2.4 Does the potential for bias remain?* Our results above showed that liberal-targeting content was systematically less accurately labelled when using individual decisions. What affect does aggregation have on this bias? In Table 7, we have re-created Table 3, but for the one-yes aggregation results.

|  | liberal-targeting | news-sharing | conservative-targeting |
|---|---|---|---|
| *Naive* | 81% | 90% | 92% |
| *Industry* | 92% | 92% | 94% |

Table 7. One-yes aggregation accuracy rates, broken down by the type of content and experimental condition.

Examining Table 7, we see two primary trends with regard to bias in accuracy, in comparison to individual decisions. First, one-yes aggregation increases accuracy for all three types of content. Second, the Industry condition seems to achieve near-parity across all types of content.

With regard to our first finding, the one-yes aggregation approach increases accuracy rates across all conditions. All six scenarios shown in Table 7 increased by 10–19% (mean = 15%) compared to individual decisions (shown in Table 3, and two increased by 16% or more. With regard to our second finding, our results suggest that for one-yes aggregation, the Industry condition may substantially diminish the disparity between liberal-targeting and conservative-targeting content. In the Industry condition, we see accuracy rates of 92% (liberal-targeting), 94% (conservative-targeting) and 92% (news-sharing). In other words, **one-yes aggregation may help mitigate the risk of unfair content labeling** that we saw in our individual labeling results above. This may be because the one-yes aggregation approach emphasizes minority perspectives that would otherwise be lost in aggregation techniques based on the majority. We return to this point in more detail in Section 5.1.1.

*4.2.5 Statistically verifying these trends.* As before, we constructed a logistic regression to evaluate our *one-yes* trends for statistically significant differences. However, because these results aggregate across multiple participants, we do not include a *participant_id* random effect. Our dependent variable was whether a one-yes decision was correct. Our independent variables were (a) the interface condition, (b) the type of post, and (c) whether the one-yes team was heterogeneous or not. We again selected the Naive-News-Sharing-Homogeneous condition as our reference category.

Table 8

|  | correct_decision |
|---|---|
| Constant | **1.5820**** (0.2261) |
| interface[*industry*] | **0.6013** . (0.2363) |
| content[*liberal-targeting*] | -0.3657 (0.2613) |
| content[*conservative-targeting*] | 0.3409 (0.3115) |
| hetergeneous[*True*] | **0.8057**** ( 0.2328) |
| Note: | . p<0.1; *p<0.05; **p<0.01 |

In Table 8 we see a significant, positive effect for heterogeneous teams. In other words, having a heterogeneous team more than doubles the odds of correctly identifying a post. We also see a suggestive positive trend for the Industry condition, which provides further evidence for the results in Table 7. Our statistical findings in Table 8 indicate no meaningful differences in the odds of correctly labeling one type of content over another, and suggest that the Industry interface condition may have a positive effect. When coupled with the results shown in Table 7, our results suggest that the one-yes aggregation technique may help ensure more equal labeling or moderation outcomes for content across the political spectrum.

## 5 DISCUSSION

### 5.1 Synthesizing Results

Overall, when considering individual content moderators, our results suggest that crowd workers perform moderately well at identifying political content, but that liberal-targeting content is less accurately labelled in comparison to conservative-targeting or news-sharing content. Thus, taking individual judgements for political labeling may produce unfair outcomes. The most common crowdsourcing aggregation technique, majority-vote also does not improve accuracy, but a one-yes aggregation technique substantially increases accuracy, and may help close the labeling gap between different kinds of political content.

*5.1.1 The importance of heterogeneity in perspectives.* First, our results suggest that for subjective, contextual settings like labeling political content, success — and fairness — in content labeling is meaningfully improved by using aggregation techniques like the *one-yes* aggregation approach that are sensitive to minority perspectives, and when aggregation teams are heterogeneous enough to reflect those minority perspectives.

While our individual labeling results do indicate the potential for unfair outcomes across different types of political content (liberal-targeting vs. conservative-targeting in our case), incorporating heterogeneous perspectives into crowd aggregation techniques may be a path towards mitigating the risk of biases in content labeling. We see this as an important direction for future work, particularly in contexts where tasks are fundamentally subjective and contextual. For instance, a common thread of research in the FAccT community has been developing definitions [2, 34, 64] of "fair" for machine learning models. Accounting for heterogeneity in context-dependent perspectives may be an important direction for ensuring fair algorithmic decision-making, moving forward.

## 5.2 Implications for Design

Beyond heterogeneity of labeling teams, interface design plays an important role as well. After all, the Industry condition in our study consistently performed best. Here, we operationalized our Industry condition using Twitter's policy for identifying political content. Recently, LinkedIn [5] has provided some guidelines about what "political" means on that platform, alongside launching a feature that allows users to exclude political content from their feeds. Our results suggest that it may be important for social media platforms to formulate and publicly communicate *opinionated* definitions of political content, and other contextual content.

*5.2.1 Capturing and utilizing political orientation.* Notably, Tables 2 and 6 suggest that an individual worker's political orientation may interact with the interface they used in a meaningful way. In our results, self-identified Republican participants (and similarly, homogeneous teams of Republicans) achieved better precision in the Naive condition, whereas the inverse was true for the Industry condition. Our findings show that some interfaces work more effectively for some workers. Our results echo and complicate results from Sen et al. [58], who found that cultural background of Turkers directly influenced their work. Developing politically- or culturally-aligned tasks for data labeling may be an important direction for future study, though we suggest that it will be important to ensure that this is not undertaken in a naive fashion. That is, carefully understanding and considering the social and ethical consequences of such designs should go hand-in-hand [e.g. 28] with pursuing this more nuanced approach to data labeling.

Practically speaking, in our study, moderators answered a two-question survey about their political views at the beginning of the task. However, there may be alternative methods, such as more thorough surveys or targeted questions, to develop a richer, more nuanced lens into content moderators' political orientations. "Just-in-time" approaches like this will also need to consider how to balance the potential value of computationally recognizing workers' and moderators' context, with their concerns about sharing personal information [56]. We see enumerating and balancing the trade-offs in privacy versus capturing crowd workers' richer context as another important direction of future work.

*5.2.2 Risks of content type bias.* Our results show that one benefit of the one-yes aggregation approach is that it may help ameliorate systematic unfairness in which political content gets recognized by a content labeling process. However, this is somewhat of a chicken-and-egg problem; recognizing the political valence of content may itself be a subjective and contextual task. In other words, this issue is precisely why human labelers are increasingly a part of content moderation pipelines to begin with. Understanding the relevant context of a given piece of content is difficult and hard to automate, particularly as the landscape of social media content continues to evolve. Therefore, it will also be important to explore reliable ways of recognizing the types of content (like political valence, in our case), and these may need to be setting-specific. Recent work in FAccT [e.g. 4] has begun to explore automated approaches for a similar problem, which may prove to be a fruitful direction over time. We see this direction as echoing the "last mile" paradox of crowdsourcing and AI as articulated by Suri and Gray [24] — that is, as needs and contexts change, human labor will always be necessary —- and there may not be robust automated ways of achieving this goal, in the general case. One solution could involve an additional step in current content moderation workflows, in which diverse crowds — along numerous dimensions [8] — parse and annotate incoming social media content to better match content with moderators.

## 5.3 Research Ethics and Social Impacts

*5.3.1 Ethical implications of this work.* A natural concern that arises around exploring the effectiveness of strategies for labeling political content is: should this work be published? After all, characterizing places where such systems are ineffective could be viewed as weaknesses for disinformation campaigns to exploit. Starbird et al. [63] showed, for instance, ways in which political disinformation campaigns exploited existing controversial topics. While we acknowledge this risk, we see building scientific knowledge about these concerns as an ethical good in and of itself. We see our work as analogous to disclosure of software vulnerabilities by computer security researchers. Making our work public and visible both provides a baseline against which platforms can conduct their own work in this space, and may also inform approaches to bias mitigation, even in the context of disinformation campaigns.

*5.3.2 Tensions in definitions of 'political'.* In our work here, we operationalized a specific definition of "political", relying on experts in relevant fields to formalize what "political" means. However, feminist [27] and other scholars have argued that "political," as a definition and a concept, can be highly personal and situated (and thus variable), so defining "political" in a systematic way may be impossible. We are sympathetic to this difficulty, while simultaneously recognizing the need for an operational definition to facilitate the pragmatic goals of labeling political content. Our work here aims to better support the practical needs of practitioners.

Moreover, above we posited that more specific, and opinionated, content moderation interfaces (and implicitly, definitions of "political content") may be an important part of achieving effective labeling in highly contextualized settings. However, making these kinds of decisions creates a new set of risks. Namely, if social media companies were to make their detailed policies for identifying political content publicly available, it may enable adversarial actors (as described by [63]) to understand the parameters and exploit these policies for their own goals. We see addressing this risk as an important direction of future work that dovetails with ongoing conversations the HCI and social computing fields broadly [20, 63] about disinformation and adversarial exploitation of social media platforms more broadly.

## 6 CONCLUSION

In this work, we took an experimental approach to studying the feasibility of content moderation to identify and label political content on social media platforms. Through a controlled experiment on the Amazon Mechanical Turk that used a real-world, ground-truth stimulus set, we found that individual content moderators perform moderately well at this task, though aggregation approaches can increase precision. We also find potential for labeling biases that advantage left-leaning political content, but crowd aggregation techniques that incorporate heterogeneity in political perspectives may help ensure more fair outcomes. We conclude by developing implications for platform designers and users, focusing on pragmatic implications of our results, and discussing a forward-looking research agenda to more deeply understand the interplay between crowd worker's political orientations and the political valence of content being labeled.

## ACKNOWLEDGMENTS

The research team would like to acknowledge the hard work of Paul Blackburn, David Mitchell, and Chris Hurt, without whom this work could not have succeeded. Partial support for this research was provided by the Office of the Vice Chancellor for Research and Graduate Education at the University of Wisconsin – Madison with funding from the Wisconsin Alumni Research Foundation. The research was partially funded by the United States National Science Foundation (NSF IIS-1651969).